\documentstyle[seceq,mbf]{ptptex}
 \def\ind{\indent}

 \def\be{\begin{equation}}
 \def\ee{\end{equation}}
 \def\ben{\begin{enumerate}}
 \def\een{\end{enumerate}}
 \def\bl{\begin{flushleft}}
 \def\el{\end{flushleft}}
 \def\bt{\begin{tabular}}
 \def\et{\end{tabular}}
 
 \def\br{\begin{flushright}}
 \def\er{\end{flushright}}
 \def\bc{\begin{center}}
 \def\ec{\end{center}}
 \def\bea{\begin{eqnarray}}
 \def\eea{\end{eqnarray}}
 \def\ba{\begin{array}}
 \def\ea{\end{array}}
 \def\bi{\begin{itemize}}
 \def\ei{\end{itemize}}

 \def\dslash{\partial{\raise 1pt\hbox{$\!\!\!/$}}}
 \def\cL{{\mbox {${\cal L}$}}}
 
 \def\cA{{\mbox {${\cal A}$}}}
 \def\cB{{\mbox {${\cal B}$}}}
 \def\cM{{\mbox {${\cal M}$}}}
 \def\cH{{\mbox {${\cal H}$}}}
 \def\cU{{\mbox {${\cal U}$}}}

 \def\cham{Chamseddine}
 \def\ind{\indent}
 \newfont{\bg}{cmr10 scaled\magstep4}
 \newcommand{\bigzerol}{\smash{\hbox{\bg 0}}}
 \newcommand{\bigzerou}{%
    \smash{\lower1.7ex\hbox{\bg 0}}}

\def\NP{Nucl.Phys.}
\def\PTP{Prog.Theor.Phys.}
\def\PL{Phys.Lett.}
\def\PR{Phys.Rev.}
\def\be{\begin{equation}}

 \def\cL{{\mbox {${\cal L}$}}}
 \def\cA{{\mbox {${\cal A}$}}}
 \def\cU{{\mbox {${\cal U}$}}}

 \def\cU{{\mbox {${\cal U}$}}}
 \def\bA{{\mbox {{\mbf A}}}}

 \def\bC{{\mbox {{\mbf C}}}}
 \def\bF{{\mbox {{\mbf F}}}}
 \def\bH{{\mbox {{\mbf H}}}}
 \def\bG{{\mbox {{\mbf G}}}}
 
 \def\bd{{\mbox {{\mbf d}}}}

\makeatletter

          \@addtoreset{equation}{section}
       \makeatother
\title{%
Gauge Theories Coupled to Fermions in Generation 
}
\vspace{5mm}
\author{%
  Hiromi {\sc Kase}, Katsusada {\sc Morita}$^{*}$ and Yoshitaka {\sc Okumura}$^{**}$
}
\inst{%
{\it  Department of Physics, 
  Daido Institute of Technology, Nagoya, 457\\
  $^{*}$Department of Physics, 
  Nagoya University, Nagoya, 464-01\\
  $^{**)}$ Department of Natural Sciences, 
  Chubu University, Kasugai, Aichi, 487}
}

\vspace{10mm}

\abst{%
Gauge theories coupled to fermions in generation
are reformulated 
in a modified version of
extended differential geometry with the symbol $\chi$.
After discussing several toy models, we will
reformulate in our framework
the standard model based on Connes' real structure.
It is shown that for the most general
bosonic lagrangin which is required 
to also reconstruct $N=2$ super Yang-Mills theory
Higgs mechanism operates
only for more than one generation as first pointed out by Connes and Lott.
}
\begin{document}
\maketitle
\section{Introduction}
An ingenious way of 'geometrizing' Higgs mechansim,
namely, unifying gauge and Higgs fields
via Dirac operator on a product space of Minkowski space-time
with two-points internal space
was proposed by
Connes\cite{1),2),3),4)} 
who
applied his noncommutative geometry (NCG)
to reconstruct the standard model.
The gauge group is defined to be the
unitary group of a noncommutative algebra
which is represented on Hilbert space of fermions.
The algebra representation (rep) nautrally leads to the concept of generation.
It is also remarkable that pure Yang-Mills action functional
in NCG automatically contains Higgs potential.
\\
\ind
NCG approach initiated by Connes
has been followed by many works\cite{5),6),7),8),9),10),11)}
in the same or similar veins.
Among them
Sitarz\cite{9)} first pointed out that NCG admits a 
differential geometric formulation
in terms of the symbol $\chi$\footnote{The symbol $\chi$ constitutes 
the fifth one-form basis
in addition to the usual four one-form basis.},
which was forced to satisfy curious properties.
For instance, unlike $dx^\mu$, $\chi$ is not closed,
$d_\chi\chi\not=0$ and the exterior product is symmetric,
$\chi\wedge\chi\not=0$.
In fact, 
it is possible to define\cite{12),13)}
a consistent noncommutative differential geometry with $d_\chi\chi=0$
for model building provided $\chi\wedge\chi\not=0$.
\\
\ind
On the other hand, it was shown by \cham\cite{14)} that 
NCG is capable to reconstruct $N=2,4$ super Yang-Mills theories.
We have notified\cite{15)}, however, that
to describe such theories using $\chi$
we have to abandon the 
symmetry $\chi\wedge\chi\not=0$
and adopt the new relation $\chi\wedge\chi=0$
with the field strength redefined.
This lets $\chi$ become more similar to the usual
one-form basis\cite{16)}.
\\
\ind
Connes\cite{4),5)} also introduced the real structure
into NCG.
This concept has been discussed
more explanatorily in several related works\cite{17),18),19)}.
It means that
flavor and color symmetries are to be implemented
in the bimodule structure of the Hilbert space.
\\
\ind
The purpose of this paper is first to
present a modified version of extended differential geometry\cite{16)},
next to apply it to broken gauge theories coupled to
fermions in generation and third to discuss the standard model in our framework
relating the real structure with the factorization property
of the standard model gauge transformations.
\\
\ind
The next section defines a modified form of
an extended differential geometry.
Toy models will be discussed
in Section 3. 
The standard model will be treated in the section 4.
The section 5 is devoted to discussions.
Some calculational details are postponed to the Appendix.
%%%%%%%%%%%%%%%%%%%%%%%%%%%%%%%%%%%%%%%%%%%%%%%%%%%%%%%%%%%%%%%%%%%%%%%%%%%%%%%
\section{ Extended differential geometry}
%%%%%%%%%%%%%%%%%%%%%%%%%%%%%%%%%%%%%%%%%%%%%%%%%%%%%%%%%%%%%%%%%%%%%%%%%%%%%%
Following Connes
we shall consider an unital, involutive noncommutative algebra \cA$\;$ together
with its $\ast$-preserving rep $\rho$ acting on the Hilbert space \cH$\;$
of fermions such that
\be
\rho(ab)=\rho(a)\rho(b),\;\rho(a+b)=\rho(a)+\rho(b),\;
\rho(1)=1,\;\rho(a^{*})=\rho(a)^{\dag}
\label{eqn:2-1}
\ee
where $a,b\in \cA$, $a\mapsto a^{*}$ is the involution
and $^{\dag}$ denotes hermitian conjugation.
In NCG $\rho(\cA)$ is the set of multiplicative operators on the fermion
Hilbert space \cH$\;$, $(a,\psi)\mapsto \rho(a)\psi, a\in\cA, \psi\in\cH$.
\\
\ind
For simplicity we omit the notation $\rho$ in the rest of this section.
The massless Dirac lagrangian takes the form
\be
\cL_D=i{\bar\psi}\gamma\cdot\partial\psi\equiv
i\langle{\tilde \psi},d\psi\rangle,\;\;%\;\;
{\tilde \psi}=\gamma_\mu\psi d{\hat x}^\mu,\;%\;\;\;
\langle d{\hat x}^\mu,d{\hat x}^\nu\rangle=g^{\mu\nu},\;%\;\;\;
g^{\mu\nu}={\rm diag}(1,-1,-1,-1)
\label{eqn:2-2}
\ee
$d$ denoting the usual
exterior derivative in ordinary differential geometry, 
$d\psi=\partial_\mu\psi d{\hat x}^\mu$, and
hat meaning dimensionless. Clearly this is not invariant under the 
gauge transformation
\be
\psi\to ^g\psi=g\psi,\;\;\;\; g\in\cU(\cA)
\label{eqn:2-3}
\ee
where the unitary group of the algebra \cA\footnote{The algebra \cA$\;$ contains
as a subset the commutative 
algebra $C^\infty(M_4)$ defined 
over Minkowski space-time $M_4$.} 
\be
\cU(\cA)=\{g\in\cA; gg^{*}=g^{*}g=1\}
\label{eqn:2-4}
\ee
defines the gauge group.
The lack of gauge invariance is remedied
by replacing 
$i\langle{\tilde \psi},d\psi\rangle$ with
\be
\cL_D=i\displaystyle{\sum_i}\langle a^{i*}{\tilde \psi},d(b^i\psi)\rangle,\;\;\;\;
a^i,b^i\in \cA
\label{eqn:2-5}
\ee
and assuming in addition to Eq.(\ref{eqn:2-3})
the following gauge transformation\footnote{We make use of the equivalence
relation $(ag^{*},g\psi)\sim(a,\psi),\;g\in\cU(\cA)$.}
\be
a^i\to ga^i, b^i\to b^ig^{*}, g\in\cU(\cA)
\label{eqn:2-6}
\ee
We assume that 
$\displaystyle{\sum_i}\langle a^{i*}{\tilde \psi},b^i\psi\rangle=
\langle {\tilde \psi},\psi\rangle$,
which implies the relation
$\displaystyle{\sum_i}a^ib^i=1$\cite{7)}\footnote{This requires a (finite)
sum in Eq.(\ref{eqn:2-5}).}.
\\
\ind
Equation (\ref{eqn:2-5}) is re-written as
\be
\cL_D=i\langle {\tilde \psi},(d+A)\psi\rangle,
\label{eqn:2-7}
\ee
where we have
introduced antihermitian, differential one-form
\be
A=\displaystyle{\sum_i}a^idb^i=A_\mu d{\hat x}^\mu,\;a^i,b^i\in\cA
\label{eqn:2-8}
\ee
Equation (\ref{eqn:2-8}) defines Yang-Mills gauge field coupled to fermions.
Because of the linearity condition
$\rho(a+b)=\rho(a)+\rho(b)$
the fermions in our theory turn out to exist in generation.
Thanks to $\displaystyle{\sum_i}a^ib^i=1$
the inhomogeneous gauge transformation 
\be
^gA=gAg^{*}+gdg^{*}
\label{eqn:2-9}
\ee
is induced by Eq.(\ref{eqn:2-6}).\\
\ind
One may now ask what happens if fermions are massive.
To generalize Eq.(\ref{eqn:2-2}) to this case
we define the extra differential to formally produce the mass term
using the chiral decomposition
of the spinor $\psi=
\normalsize{\left(
             \ba{cl}
             \psi_L\\
             \psi_R\\
             \ea
             \right)}\in\cH$
by
\be
d_\chi\psi=M\psi\chi, \;M=\left(
                          \ba{cc}
                          0&M_1\\
                          M_1^{\dag}&0\\
                          \ea
                          \right)=M^{\dag}
\label{eqn:2-10}
\ee
where $M$ is fermion mass matrix.
The extra one-form basis $\chi$ combines with the usual ones
$\{d{\hat x}^\mu\}_{\mu=0,1,2,3}$
to form 5-dimensional one-form basis:
\be
d{\hat x}^\mu\wedge d{\hat x}^\nu=-d{\hat x}^\nu\wedge d{\hat x}^\mu,\;
d{\hat x}^\mu\wedge \chi=-\chi\wedge d{\hat x}^\mu,\;
\chi\wedge \chi=0
\label{eqn:2-11}
\ee
Assuming further
\be
\langle d{\hat x}^\mu,\chi\rangle=\langle\chi,d{\hat x}^\mu\rangle=0,\;\;\;\;
\langle \chi,\chi\rangle=-1
\label{eqn:2-12}
\ee
we obtain the Dirac lagrangian for massive fermion through
\be
\cL_D=i\langle{\tilde \psi},\bd\psi\rangle,\;\;\;\;
{\tilde \psi}=\gamma_\mu\psi d{\hat x}^\mu+i\psi\chi
\label{eqn:2-13}
\ee
where $\bd=d+d_\chi$. As in the massless case Eq.(\ref{eqn:2-13})
is not invariant under Eq.(\ref{eqn:2-3}).
But the recipe is now clear.
The sum
\be
\cL_D=i\displaystyle{\sum_i}\langle a^{i*}{\tilde \psi},\bd(b^i\psi)\rangle,\;\;\;\;
a^i,b^i\in \cA
\label{eqn:2-14}
\ee
becomes gauge invariant under Eqs.(\ref{eqn:2-3}) and (\ref{eqn:2-6}).
Putting
\be
\bA=\displaystyle{\sum_i}a^i\bd b^i=A_\mu d{\hat x}^\mu+\Phi\chi
\label{eqn:2-15}
\ee
we rewrite Eq.(\ref{eqn:2-14}) as
\be
\cL_D=i\langle {\tilde \psi},(\bd+\bA)\psi\rangle
\label{eqn:2-16}
\ee
Consequently,
Eq.(\ref{eqn:2-14}) introduces gauge-invariant
coupling of scalars, $\Phi$, $\chi$ being assumed
to be Lorentz scalar,
to fermions.\\
\ind
The gauge transformation property of \bA$\;$
under Eq.(\ref{eqn:2-6})
determines that of $\Phi$ in addition to Eq.(\ref{eqn:2-9}):
\be
^g\Phi\chi=g\Phi g^{*}\chi+gd_\chi g^{*}
\label{eqn:2-17}
\ee
where use has been made of the relation
$\displaystyle{\sum_i}a^ib^i=1$.
Unless the inhomogeneous term in Eq.(\ref{eqn:2-17})
vanishes $\Phi$ is not a physical field.
\\
\ind
In writing Eqs.(\ref{eqn:2-14}), (\ref{eqn:2-15})
and (\ref{eqn:2-16}) we have implicitly assumed the Leipniz rule
\be
d_\chi(f\psi)=(d_\chi f)\psi+f(d_\chi\psi),\;\;f\in\cA,\;\;\chi\psi=\psi\chi
\label{eqn:2-18}
\ee
Note that with the chiral decomposition of the spinor
we must write
elements of \cA$\;$ as 2$\times$2 matrix in block form like
$
f =\normalsize{\left(
          \ba{cc}
          f_1&g_1\\
          g_2&f_2\\
          \ea
          \right)}\in\cA
$.
It follows that\footnote{In the previous papers\cite{12),13),16)}
we derived or assumed $d_\chi M\not=0$. The present modification 
seems to be more natural.}
\be
d_\chi f=(Mf-fM)\chi,\;\;f\in\cA,\;\;d_\chi M=0
\label{eqn:2-19}
\ee
which in turn implies the following Leipniz rule
\be
d_\chi(fg)=(d_\chi f)g+f(d_\chi g),f,g\in\cA,\;\;\chi g=g\chi
\label{eqn:2-20}
\ee
The extended 
operator $\bd$ is nilpotent
\be
\bd^2=0
\label{eqn:2-21}
\ee
since $d_\chi\chi=0$\cite{12)} and $d^2=d_\chi^2=dd_\chi+ d_\chi d=0$.
We need to prove the last
equality:
\be\ba{cl}
&(dd_\chi+d_\chi d)f=d(Mf-fM)\chi+d_\chi df
=(M\partial_\mu f-\partial_\mu fM)(d{\hat x}^\mu\wedge\chi
+\chi\wedge d{\hat x}^\mu)=0\\[2mm]
&(dd_\chi+d_\chi d)\psi=d(M\psi)\chi+d_\chi (d\psi)
=M\partial_\mu \psi(d{\hat x}^\mu\wedge\chi
+\chi\wedge d{\hat x}^\mu)=0
\label{eqn:2-22}
\ea\ee
\ind
According to Eq.(\ref{eqn:2-19}) the inhomogeneous term in 
the right-hand side of Eq.(\ref{eqn:2-17})
vanishes if the mass term is gauge invariant, $[M,g]=0$.
Then $\Phi$ is a physical field irrelevant to generation of fermion masses.
As an example one may remark that, in this case,
there is no need to make the sum (\ref{eqn:2-14}),
but 
\be
\cL_D=i\displaystyle{\sum_i}\langle a^{i*}{\tilde \psi},d(b^i\psi)\rangle+
i\langle {\tilde \psi},d_\chi\psi\rangle,\;\;\;\;
a^i,b^i\in \cA
\label{eqn:2-23}
\ee
is already gauge invariant.
Hence we are led back to simply add gauge invariant mass term
to Eq.(\ref{eqn:2-7}), leading to no scalars coupled to fermions.
\\
\ind
The decomposition (\ref{eqn:2-15}) of the generalized one-form $\bA$
into gauge $A_\mu$ and shifted Higgs $\Phi$ fields corresponds to 
Connes' prescription of NCG to unify gauge and Higgs fields,
where the mass term in the Dirac operator yields the
shifted Higgs field.
Likewise, our $d_\chi$ gives rise to $\Phi$.
\\
\ind
We now go on to define the field strength corresponding to \bA.
It will yield the bosonic lagrangian
involving solely the bosons coupled to fermions.
We employ the definition by Clifford product
as proposed in Ref.16).
\be\ba{cl}
\bG&=\bd\vee\bA+\bA\vee\bA=\bF+\bF_0\\[2mm]
\bF&=\bd\wedge\bA+\bA\wedge\bA\equiv \bd\bA+\bA^2,\;\;
\bd\bA=\displaystyle{\sum_i}\bd a^i\wedge\bd b^i \\[2mm]
\bF_0&=\langle \bd,\bA\rangle+\langle \bA,\bA\rangle,\;
\langle \bd,\bA\rangle=\displaystyle{\sum_i}\langle \bd a^i,
\bd b^i\rangle 
\label{eqn:2-24}
\ea\ee
The field strength \bG$\;$ is inhomogeneous in the rank of differential forms.
In other words,
it is given by the sum of two-form \bF$\;$ and zero-form $\bF_0$\footnote{Similarly,
Connes' field strength consists of three terms,
$\theta=\theta_0+\theta_1+\theta_2$ corresponding to two-form $\theta_0\propto
[\gamma^\mu,\gamma^\nu]$, three form $\theta_1\propto\gamma^\mu\gamma_5$
and zero-form $\theta_2$ containing no Dirac matrices, $\gamma^\mu$ being regarded as
one-form basis.}.
Both \bF$\;$ and $\bF_0$ are gauge covariant so is \bG
\footnote{The fact that both \bF$\;$ and $\bF_0$ are separately gauge
covariant implies that the most general field strength in this scheme
contains an arbitrary real parameter $\kappa$:
$\bG=\bF+\kappa\bF_0$.Then the arbitrary parameter $\kappa^2$ 
would appear in $V$. For simplicity
in this paper 
we put $\kappa=1$ so that the field strength is given by the Clifford product.}.
Explicitly we have
\be\ba{cl}
\bF&=F+DH\wedge\chi,\;F=dA+A^2,\;DH=dH+AH-HA,\;H=\Phi+M\\[2mm]
\bF_0&=X-(H^2-M^2-Y),\;X=-\displaystyle{\sum_i}a^i\partial^2b^i+\partial^\mu A_\mu
+A^\mu A_\mu=\left(
             \ba{cc}
             X_1&0\\
             0&X_2\\
             \ea
             \right)\\[2mm]
A_\mu&=\displaystyle{\sum_i}a^i\partial_\mu b^i,\;\Phi=
\displaystyle{\sum_i}a^i[M,b^i],\;
Y=\displaystyle{\sum_i}a^i[M^2,b^i]=\left(
                                    \ba{cc}
                                    Y_1&0\\
                                    0&Y_2\\
                                    \ea
                                    \right)
\label{eqn:2-25}
\ea\ee
\ind
The system is governed by the lagrangian
\be
\cL=\cL_D+\cL_B
\label{eqn:2-26}
\ee
where
the fermionic sector is given by Eq.(\ref{eqn:2-16}), while
the bosonic sector is obtained by
\be
\cL_B=-\displaystyle{{1\over 4}}\langle\!\langle z^2\bG,\bG\rangle\!\rangle
=-\displaystyle{{1\over 4}}\langle\!\langle z^2\bF,\bF\rangle\!\rangle-V,\;\;
V=\displaystyle{{1\over 4}}\langle\!\langle z^2\bF_0,\bF_0\rangle\!\rangle
\label{eqn:2-27}
\ee
where $z^2$ is a positive
matrix commuting with the gauge group and
we have used the fact that two-form and zero-form are orthogonal.\footnote{In 
evaluating Eq.(\ref{eqn:2-27}) we use the inner product
$$
\langle d{\hat x}^\mu\wedge d{\hat x}^\nu, d{\hat x}^\rho\wedge d{\hat x}^\sigma\rangle=
g^{\mu\rho}g^{\nu\sigma}-g^{\mu\sigma}g^{\nu\rho},\;
\langle d{\hat x}^\mu\wedge \chi, d{\hat x}^\nu\wedge\chi\rangle=-g^{\mu\nu}
$$
}
\\
\ind
The lagrangian
depends on the model. In the next section we
shall discuss toy models.
%%%%%%%%%%%%%%%%%%%%%%%%%%%%%%%%%%%%%%%%%%%%%%%%%%%%%%%%%%%%%%%%%%%%%%%%%%%%%%%
\section{Toy models}%Section3
%%%%%%%%%%%%%%%%%%%%%%%%%%%%%%%%%%%%%%%%%%%%%%%%%%%%%%%%%%%%%%%%%%%%%%%%%%%%%%
First of all we shall consider
$SU(2)$ gauge theory.
As is well known there exists an infinite tower of unitary irreps of $SU(2)$
with dimensions $2j+1, j=0,{1\over 2},1,\cdots$. $SU(2)$ gauge theory
is defined for the spinor belonging to any
one of them.\\
\ind 
Now local $SU(2)$ is defined by the unitaries of the algebra
$\cA=C^\infty(M_4)\otimes \bH$.
Real quaternion \bH$\;$ has only one irrep
\be
\bH\ni a\mapsto\left(
                     \ba{cc}
                     \alpha&\beta\\
                     -\beta^{*}&\alpha^{*}\\
                     \ea
                     \right),\;\alpha,\beta\in\bC
\label{eqn:3-1}
\ee
Hence one can write $SU(2)=\{g\in\bH; gg^{*}=g^{*}g=1,g^{*}={\bar g}\}$,
where ${\bar g}$ is quaternion conjugation. 
Using the irrep (\ref{eqn:3-1}) \bH$\;$ is represented by
\be
\bH\ni a\mapsto \rho(a)={\rm diag}(a,\cdots,a^{*},\cdots),\;a^{*}={\rm c.c.}\;{\rm of}
\;a
\label{eqn:3-2}
\ee
which acts on a repetition of fundamental reps of $SU(2)$.
Consequently, 
fermions exist in generation and
must be non-chiral so that $M$ commutes with $SU(2)$ and $\Phi=
\displaystyle{\sum_i}a^i[M,b^i]=0$. 
Thus $H=M$ is constant, commuting with $A$,
hence $DH=0$. We then conclude that $\cL_B$ contains only Yang-Mills
term because $Y=0$ and $V$ vanishes due to the equation of motion.
That is, our algebraic recipe leads to $SU(2)$ gauge theory without scalars,
where fermion doublets exist in generation.
\\
\ind
Higgs mechanism can be incorporated into our scheme by enlarging the algebra to
$\cA=C^\infty(M_4)\otimes(\bH\oplus\bC)$.
The unitary group is Map($M_4,SU(2)\times U(1))=
\cU(C^\infty(M_4)\otimes(\bH\oplus\bC))$.
Writing
\be
\bA=\displaystyle{\sum_i}\rho(a_1^i,b_1^i)\bd \rho(a_2^i,b_2^i)
\label{eqn:3-3}
\ee
for Eq.(\ref{eqn:2-15})
we assume the following rep of \cA$\;$ on the spinor with chiral components
\be
\rho(a,b)=\left(
          \ba{cc}
          a\otimes 1_{N_g}&0\\
          0&b\otimes 1_{N_g}\\
          \ea
          \right),\;a\in\bH,\;b\in\bC
\label{eqn:3-4}
\ee
so that
left-handed and right-handed spinors belong to doublet and singlet, 
respectively\footnote{There is no reason to assume the same $N_g$
in the doublet and singlet sectors. This assumption is only for later
convenience. The $N\times N$ unit matrix is denoted by $1_N$.}.
Except for generation indices
$\psi_L$ has two-components like $\left(
                                  \ba{cl}
                                  \nu_L\\
                                  e_L\\
                                  \ea
                                  \right)$, while
$\psi_R$ has only one component like $e_R$.
Here and hereafter we omit infinite-dimentional part $C^\infty(M_4)$ of the algebra
and remember that, say, $a$ in Eq.(\ref{eqn:3-4}) is \bH-valued local functions
represented by Eq.(\ref{eqn:3-1}).
Choosing the mass matrix as
\be
M=\left(
  \ba{cc}
  0&M_1\\
  M_1^{\dag}&0\\
  \ea
  \right),\;\;M_1=\left(
                  \ba{cl}
                  0\\
                  m\\
                  \ea
                  \right)
\label{eqn:3-5}
\ee
where $m$ is $N_g\times N_g$ mass matrix,
we find from the definition of $\Phi$ in Eq.(\ref{eqn:2-25})
\be
\Phi=\left(
     \ba{cc}
     0&\varphi M_1\\
     M_1^{\dag}\varphi^{\dag}&0\\
    \ea
    \right),\;\;\varphi=\left(
                  \ba{cc}
                  \varphi_0^{*}&\varphi_+\\
                  -\varphi_-&\varphi_0\\
                  \ea
                  \right)
\label{eqn:3-6}
\ee                  
Note that $\varphi$ is \bH-valued scalar function.
Also
\be\ba{cl}
&H=\Phi+M=\left(
         \ba{cc}
         0&hM_1\\
         M_1^{\dag}h^{\dag}&0\\
         \ea
         \right)\\[2mm]
&h=\varphi+1_2=\left(
             \ba{cc}
             \phi_0^{*}&\phi_+\\
             -\phi_-&\phi_0\\
             \ea
             \right),\;\;\phi_0=\varphi_0+1,\;\; \phi_{\pm}=\varphi_{\pm}
\label{eqn:3-7}
\ea\ee
\ind
It can be shown that Eq.(\ref{eqn:2-16}) leads to
\be\ba{cl}
\cL_D&={\bar\psi}_Li\gamma^\mu(\partial_\mu-\displaystyle{{ig_2\over 2}}
A_\mu^{(2)})\psi_L+{\bar\psi}_Ri\gamma^\mu(\partial_\mu-\displaystyle{{ig_1\over 2}}
A_\mu^{(1)})\psi_R\\[2mm]
&\;\;-{\bar\psi}_Lm\phi\psi_R-{\bar\psi}_R\phi^{\dag}m^{\dag}\psi_L
\label{eqn:3-8}
\ea\ee
where
\be
\phi=\left(
     \ba{cl}
     \phi_+\\
     \phi_0\\
     \ea
     \right),\;\;\langle\phi\rangle=\left(
                                    \ba{cl}
                                    0\\
                                    1\\
                                    \ea
                                    \right)
\label{eqn:3-9}
\ee
is the normalized Higgs field
and we put
\be
A=\displaystyle{\sum_i}\rho(a_1^i,b_1^i)d \rho(a_2^i,b_2^i)
=\left(
 \ba{cc}
 -\displaystyle{{ig_2\over 2}}A_\mu^{(2)}d{\hat x}^\mu\otimes 1_{N_g}&0\\
 0&-\displaystyle{{ig_1\over 2}}A_\mu^{(1)}d{\hat x}^\mu\otimes 1_{N_g}\\
 \ea
 \right)
\label{eqn:3-10}
\ee
$A_\mu^{(2)}$ is $SU(2)$ gauge field,
while  $A_\mu^{(1)}$  $U(1)$ gauge field, both hermitian,
$g_2$ and $g_1$ being the corresponding gauge coupling constants, respectively.\\
\ind
Let us now evaluate the bosonic lagrangian (\ref{eqn:2-27}) using the formulae
(\ref{eqn:2-25}). Putting
\be
z^2=\displaystyle{{4\over N_g}}
        \left(
        \ba{cc}
        g_2^{-2}1_2\otimes 1_{N_g}&0\\
        0&2g_1^{-2}\otimes 1_{N_g}\\
        \ea
        \right)
\label{eqn:3-11}
\ee
we have for the two-form piece
\be
-\displaystyle{{1\over 4}}\langle\!\langle z^2\bF,\bF\rangle\!\rangle
=-\displaystyle{{1\over 8}}{\rm tr}F_{\mu\nu}^{(2)}F^{(2)\mu\nu}
-\displaystyle{{1\over 4}}F_{\mu\nu}^{(1)}F^{(1)\mu\nu}+L(D^\mu\phi)^{\dag}
(D_\mu\phi)
\label{eqn:3-12}
\ee
where $F_{\mu\nu}^{(i)}$ are field strength of the gauge fields
$A_\mu^{(i)}, i=1,2, D_\mu\phi=(\displaystyle{\partial_\mu}
-\displaystyle{{ig_2\over 2}}A_\mu^{(2)}
+\displaystyle{{ig_1\over 2}}A_\mu^{(1)})\phi$ and we define
\be
L=\displaystyle{{2\over N_g}}(\displaystyle{{1\over g_2^2}}+
\displaystyle{{2\over g_1^2}}){\rm tr}_g(m^{\dag}m)
\label{eqn:3-13}
\ee
with tr$_g$ denoting the trace in the generation space.
Next we compute the zero form piece.
Since $Y_2=0$ in Eq.(\ref{eqn:2-25}),
$X_1,X_2$ and $Y_1$ are auxiliary fields to be eliminated from the lagrangian
using the equation of motion, we get
\be
V=K(\phi^{\dag}\phi-1)^2,\;\;\;
K=\displaystyle{{2\over N_gg_1^2}}({\rm tr}_g(M_1^{\dag}M_1)^2
-\displaystyle{{1\over N_g}}({\rm tr}_gM_1^{\dag}M_1)^2)
\label{eqn:3-14}
\ee
Finally, we have
\be
\cL_B=-\displaystyle{{1\over 8}}{\rm tr}F_{\mu\nu}^{(2)}F^{(2)\mu\nu}
-\displaystyle{{1\over 4}}F_{\mu\nu}^{(1)}F^{(1)\mu\nu}+(D^\mu\phi)^{\dag}(D_\mu\phi)
-\displaystyle{{\lambda\over 4}}
(\phi^{\dag}\phi-\displaystyle{{v^2\over 2}})^2
\label{eqn:3-15}
\ee
where
\be
\lambda=\displaystyle{{4K\over L^2}},\;\;
v^2=2L
\label{eqn:3-16}
\ee
For $\lambda$ should be positive, the theory is consistently defined only if
$N_g>1$.\\
\ind
The rescaling $\phi\to{1\over\sqrt{L}}\phi={\sqrt{2}\over v}\phi$ to obtain
above $\cL_B$ renders the mass term in Eq.(\ref{eqn:3-8}) multiplied by
${\sqrt{2}\over v}$.\\
\ind
There is anther rep of the same algebra other than (\ref{eqn:3-4}):
\be
\rho(a,b)=\left(
          \ba{cc}
          a\otimes 1_{N_g}&0\\
          0&B\otimes 1_{N_g}\\
          \ea
          \right),\;\;B=\left(
                        \ba{cc}
                        b&0\\
                        0&b^{*}\\
                        \ea
                        \right)
\label{eqn:3-17}
\ee
which is suitable for 'massive' neutrino. This rep acts on doublet like
$\psi_L=\left(
        \ba{cl}
        \nu\\
        e\\
        \ea
        \right)_L$ and
singlets like
$\psi_R=\left(
        \ba{cl}
        \nu_R\\
        e_R\\
        \ea
        \right)$ with generation indices omitted.
The mass matrix is given by
\be
M_1=\left(
    \ba{cc}
    m_1&0\\
    0&m_2\\
    \ea
    \right),\;\;\;\;m_{1,2}:\;N_g\times N_g
\label{eqn:3-18}
\ee
The Higgs field is given by
Eqs.(\ref{eqn:3-6}) and (\ref{eqn:3-7}) with $M_1$ being of the form (\ref{eqn:3-18}).
Putting
\be\ba{cl}
A&=\displaystyle{\sum_i}\rho(a_1^i,b_1^i)d \rho(a_2^i,b_2^i)\\[2mm]
&
=\left(
 \ba{cc}
 -\displaystyle{{ig_2\over 2}}A_\mu^{(2)}d{\hat x}^\mu\otimes 1_{N_g}&0\\
 0&-\displaystyle{{ig_1\over 2}}\left(
                                \ba{cc}
                                1&0\\
                                0&-1\\
                                \ea
                                \right)A_\mu^{(1)}d{\hat x}^\mu\otimes 1_{N_g}\\
 \ea
 \right)
\label{eqn:3-19}
\ea
\ee
we find
\be\ba{cl}
\cL_D&={\bar\psi}_Li\gamma^\mu(\partial_\mu-\displaystyle{{ig_2\over 2}}
A_\mu^{(2)})\psi_L+{\bar\psi}_Ri\gamma^\mu(\partial_\mu-\displaystyle{{ig_1\over 2}}
\left(
\ba{cc}
1&0\\
0&-1\\
\ea
\right)
A_\mu^{(1)})\psi_R\\[2mm]
&\;\;-{\bar\psi}_L(m_1{\tilde\phi},m_2\phi)\psi_R
-{\bar\psi}_R\left(
             \ba{cl}
             {\tilde\phi}^{\dag}m_1^{\dag}\\
             \phi^{\dag}m_2\\
             \ea
             \right)
             \psi_L,\;{\tilde \phi}=\left(
                                    \ba{cc}
                                    0&1\\
                                    -1&0\\
                                    \ea
                                    \right)\phi^{*}
\label{eqn:3-20}
\ea\ee
The bosonic lagrangian takes the same form as (\ref{eqn:3-15})
with $L$ of Eq.(\ref{eqn:3-16}) given by
\be
L=\displaystyle{{2\over N_g}}(\displaystyle{{1\over g_2^2}}+
\displaystyle{{1\over g_1^2}}){\rm tr}_g(M_1^{\dag}M_1)
\label{eqn:3-21}
\ee
provided we set
\be
z^2=\displaystyle{{4\over N_g}}
        \left(
        \ba{cc}
        g_2^{-2}1_2\otimes 1_{N_g}&0\\
        0&g_1^{-2}1_2\otimes 1_{N_g}\\
        \ea
        \right)
\label{eqn:3-22}
\ee
\ind
Next consider the case $N_g$=1 and let $\psi^a (a=1,2,3)$ 
belong to the adjoint rep of $SU(2)$ 
with the matrix field
\be
\psi=\tau_a\psi^a
\label{eqn:3-23}
\ee
where ${\mbf\tau}=(\tau_1,\tau_2,\tau_3)$ is the Pauli matrix.
Choosing the algebra $\cA=C^\infty(M_4)\otimes M_2(\bC)$
whose unitary group is Map($M_4,SU(2)\times U(1))
=\cU(C^\infty(M_4)\otimes M_2(\bC))$, although $U(1)$ factor is
automatically eliminated in the reconstruction below,
we represent it by
\be
\rho(a)=\left(
          \ba{cc}
          a&0\\
          0&a\\
          \ea
          \right),\;a\in M_2(\bC)
\label{eqn:3-24}
\ee
Action by $\bd+\bA$ on $\psi$ is defined by the commutator
\be\ba{cl}
(\bd+\bA)\psi&\equiv [\bd+\bA,\psi]=[d+A,\psi]+[d_\chi+\Phi\chi,\psi]\\[2mm]
[d+A,\psi]&\equiv d\psi+\left(
                        \ba{cl}
                        [A_1,\psi_L]\\[1mm]
                        [A_2,\psi_R]\\[1mm]
                        \ea
                        \right), A=\left(
                                   \ba{cc}
                                   A_1&0\\
                                   0&A_2\\
                                   \ea
                                   \right)\\[2mm]
[d_\chi,\psi]&=[M,\psi]\equiv \left(
                             \ba{cl}
                             [M_1,\psi_R]\\[1mm]
                             [M_1^{\dag},\psi_L]\\[1mm]
                             \ea
                             \right),\;\;
[\Phi,\psi]\equiv \left(
                   \ba{cl}
                   [\varphi,\psi_R]\\[1mm]
                   [\varphi^{\dag},\psi_L]\\[1mm]
                   \ea
                   \right)
\label{eqn:3-25}
\ea\ee
where
\be
\Phi=\displaystyle{\sum_i}
     \rho(a^i)[M,\rho(b^i)]
    =\left(
     \ba{cc}
     0&\varphi\\
     \varphi^{\dag}&0\\
     \ea
     \right),\;\;a^i,b^i\in\cA
\label{eqn:3-26}
\ee
By assumption $A_1=A_2$ and we put $A_{1\mu}=-igA_\mu$
where $A_\mu$ is hermitian $SU(2)$ gauge field due to the commutator
(\ref{eqn:3-25}).
Hence the gauge group is $SU(2)$, not $SU(2)\times U(1)$.
Another comment here is that $\Phi$ does not take the form (\ref{eqn:3-6})
but has three complex components in view of the commutator (\ref{eqn:3-25}):
\be
\varphi
    =\left(
     \ba{cc}
     \varphi_{11}&\varphi_{12}\\
     \varphi_{21}&-\varphi_{11}\\
     \ea
     \right)
\label{eqn:3-27}
\ee
\\
\ind
$H=\Phi+M$ is given by Eq.(\ref{eqn:3-27})
with $\varphi\to h=\varphi+M_1$, whence we put
\be
-\displaystyle{{1\over g}}h
     \equiv\phi=\tau_a\phi^a\equiv S-iP
\label{eqn:3-28}
\ee
It is now easy to evaluate\cite{15)} the lagrangian
\be
\cL_B=-\displaystyle{{1\over 4g^2}}\langle\!\langle \bG,\bG\rangle\!\rangle
=-\displaystyle{{1\over 4g^2}}\langle\!\langle \bF,\bF\rangle\!\rangle-V,\;\;
V=\displaystyle{{1\over 4g^2}}\langle\!\langle \bF_0,\bF_0\rangle\!\rangle
\label{eqn:3-29}
\ee
where we put $z^2=\displaystyle{{1\over g^2}}1_4$ in Eq.(\ref{eqn:2-27})
to take into account supersymmerty. The result is
\be
\cL_B=-\displaystyle{{1\over 8}}{\rm tr}F_{\mu\nu}F^{\mu\nu}
+\displaystyle{{1\over 2}}{\rm tr}([D^\mu,\phi])^{\dag}[D_\mu,\phi])
-\displaystyle{{g^2\over 8}}{\rm tr}[\phi,\phi^{\dag}]^2
\label{eqn:3-30}
\ee
where $D_\mu=\partial_\mu-igA_\mu$, provided
$M_1$ is a normal matrix,
$M_1{M_1}^{\dag}={M_1}^{\dag}M_1$\footnote{This implies that 
$Y=0$ in Eq.(\ref{eqn:2-25}). Hence there must be $X$ term in
$\bF_0$ to obtain the scalar potential.}.\\
\ind
In the fermionic sector we simply find from Eqs.(\ref{eqn:2-16}),
(\ref{eqn:3-25}), (\ref{eqn:3-26}) and (\ref{eqn:3-28})
\be
\cL_D={\rm tr}({\bar \psi}i\gamma^\mu [D_\mu,\psi]
+g{\bar\psi}[S+i\gamma_5P,\psi])
\label{eqn:3-31}
\ee
As is well known the sum of Eqs.(\ref{eqn:3-30}) and
(\ref{eqn:3-31}) defines $N=2$ super Yang-Mills theory\cite{20)}.
%%%%%%%%%%%%%%%%%%%%%%%%%%%%%%%%%%%%%%%%%%%%%%%%%%%%%%%%%%%%%%%%%%%%%%%%%%%%%%%
\section{Standard model of elementary particles}%Section4
%%%%%%%%%%%%%%%%%%%%%%%%%%%%%%%%%%%%%%%%%%%%%%%%%%%%%%%%%%%%%%%%%%%%%%%%%%%%%%
\ind
To reconstruct the standard model of elementary
particles in the present scheme 
we take into account 
Connes real structure $J$. To define $J$ as simple as possible
we follow Connes to double spinors using charge conjugate spinor $\psi^c$ as
\be
\Psi=\left(
     \ba{cl}
     \psi\\
     \psi^c\\
     \ea
     \right),\; \psi=\left(
     \ba{cl}
     \psi_L\\
     \psi_R\\
     \ea
     \right)
\label{eqn:4-1}
\ee
upon which gauge transformation acts like
\be
^g\Psi=U\Psi=U_1U_2\Psi,\;\;U^{\dag}U={U_i}^{\dag}U_i=1
\label{eqn:4-2}
\ee
if
\be
U=\left(
  \ba{cc}
  u_1{u_2}^{*}&0\\
    0&{u_1}^{*}u_2\\
    \ea
    \right),\;\;\;\;{u_i}^{*}={\rm c.c.}\;{\rm of}\;u_i
\label{eqn:4-3}
\ee
where 
\be
U_1=\left(
    \ba{cc}
    u_1&0\\
    0&u_2\\
    \ea
    \right),\;\;\;\;\;
U_2=\left(
    \ba{cc}
    {u_2}^{*}&0\\
    0&{u_1}^{*}\\
    \ea
    \right)
\label{eqn:4-4}
\ee
with ${u_1}^{*}u_2=u_2{u_1}^{*}$, or $U_1U_2=U_2U_1$. 
In the language of Connes mathematics obtaining $U_2$
from $U_1$ is accomplished by means of the real structure $J$, $U_2=JU_1J^{\dag}$, 
provided 
$U_1$ is a rep
of a noncommutative algebra\footnote{The operator $J$
is an anti-linear isometry in the total Hilbert space \cH$\;$
which satisfies two conditions\cite{4),5)}. Then Eq.(\ref{eqn:4-2})
reads $^g\Psi= U_1\Psi U_1^{*}\equiv U_1JU_1J^{*}\Psi$.
}. 
Since Abelian charges cancel among fermions and anti-fermions,
det$U$=1. This condition is satisfied if
\be
{\rm det}u_1{u_2}^{*}=1
\label{eqn:4-5}
\ee
which is highly nontrivial.
We shall refer Eq.(\ref{eqn:4-5})
to as Connes unimodularity condition.\\
\ind
In the standard model we choose the basis using the mass eigenstates
\be
%\psi^T=(q_L,l_L,u_R,d_R,\nu_{eR},e_R),\;\;
\psi=\left(
     \ba{cl}
     q_L\\
     l_L\\
     u_R\\
     d_R\\
     \nu_{eR}\\
     e_R\\
     \ea
     \right),\;\;\;\; 
     q_L=\left(
         \ba{cl}
          u\\
         U_qd\\
         \ea
         \right)_L,\;\;%\;\;
     l_L=\left(
         \ba{cl}
         \nu_e\\
         U_le\\
         \ea
         \right)_L
\label{eqn:4-6}
\ee     
where we omit generation indices on leptons and quarks
as well as color indices on quarks and similarly for anti-fermions.
The $N_g\times N_g$ Kobayashi-Maskawa matrices are denoted by $U_{l,q}$
in the lepton and quark sectors, respectively.
Experiment shows $N_g=3$ up to the present energy
but we let $N_g$ be a free parameter.
We assume massive neutrinos using the rep (\ref{eqn:3-17}) in the lepton sector.\\
\ind
To recover the standard gauge group $SU(3)\times SU(2)_L\times U(1)_Y$
from $\cU(\cA)$
we make Connes choice\cite{4),5)}
\be
\cA=C^\infty(M_4)\otimes(\bH\oplus\bC\oplus M_3(\bC))
\label{eqn:4-7}
\ee     
whose unitaries are Map($M_4,U(3)\times SU(2)\times U(1)$).
We shall see below that the unimodularity condition (\ref{eqn:4-5}) reduces
$U(3)\times SU(2)\times U(1)$ to $SU(3)\times SU(2)_L\times U(1)_Y$
with correct hypercharge assignment.\\
\ind
In the basis (\ref{eqn:4-6}) Eqs.(\ref{eqn:4-2}), (\ref{eqn:4-3}) 
and (\ref{eqn:4-4})
turn out to be given by the unitary restriction of the rep of
Eq.(\ref{eqn:4-7}): $u_1=\rho_w(a,b)$ and  $u_2=\rho_s(b,c)$ for
$(a,b,c)\in\cU(\cA)$ where, for general $a\in\bH, b\in\bC, c\in M_3(\bC)$,
\cA$\;$ is represented on Eq.(\ref{eqn:4-1}) by\cite{4),5),17),18),19)}
\be\ba{cl}
\rho(a,b,c)&=\left(
    \ba{cc}
    \rho_w(a,b)&0\\
    0&\rho_s(b,c)\\
    \ea
    \right)\\[2mm]
\rho_w(a,b)&=\left(
    \ba{cc}
    \rho_1(a)&0\\
    0&\rho_2(b)\\
    \ea
    \right),\;\;\;
    \rho_s(b,c)=\left(
    \ba{cc}
    \rho_3(b,c)&0\\
    0&\rho_4(b,c)\\
    \ea
    \right)\\[2mm]
&\rho_1(a)=\left(
           \ba{cc}
           a\otimes 1_3\otimes 1_{N_g}&0\\
           0&a\otimes 1_{N_g}\\
           \ea
           \right),\;\;
\rho_2(b)=\rho_1(B)\\[2mm]
&\rho_3(b,c)=\left(
             \ba{cc}
             1_2\otimes c^{*}\otimes 1_{N_g}&0\\
             0&b1_2\otimes 1_{N_g}\\
             \ea
             \right)=\rho_4(b,c)
\label{eqn:4-8}
\ea\ee     
where $B$ is given by Eq.(\ref{eqn:3-17}).
By $U_1$ flavor acts only 
in the particle sector,
whereas color operates only in the
anti-particle sector\footnote{Before introduction of the real structure,
Connes considered\cite{3)} the bimodule $\cA\oplus\cB,\;
\cA=\bH\oplus\bC,\;\cB=\bC\oplus M_3(\bC)$ with left action
by flavor and right action by color. Thanks to the real structure $J$,
the bimodule structure of \cH$\;$ is obtained with single
algebra (\ref{eqn:4-7}) just because 
there exist matter and anti-matter in quantum field theory.}. 
\\
\ind
The well-known fermionic sector of the standard model lagrangian
is given using the doubled spinor (\ref{eqn:4-1}) by
\be
\cL_D=\displaystyle{{i\over 2}}\langle{\tilde\Psi},(\bd+\bA+J\bA J^{\dag})\Psi\rangle
\label{eqn:4-9}
\ee     
where 
\be\ba{cl}
d_\chi\Psi&=\cM\Psi\chi,\;\;\cM=\cM_1+J\cM_1 J^{\dag},\;\;\;
                      \cM_1=\left(
                           \ba{cc}
                           M&0\\
                           0&0\\
                           \ea
                           \right),\;\;
                           M=\left(
                           \ba{cc}
                           0&M_1\\
                           {M_1}^{\dag}&0\\
                           \ea
                           \right)\\[2mm]
\bA&=\displaystyle{\sum_i}\rho(a_1^i,b_1^i,c_1^i)\bd\rho(a_2^i,b_2^i,c_2^i)
    =A+\Phi\chi\\[2mm]
A&=\left(
    \ba{cc}
    A_w&0\\
    0&A_s\\
    \ea
    \right),\;\;
    A_w=\displaystyle{\sum_i}\rho_w(a_1^i,b_1^i)d\rho_w(a_2^i,b_2^i),\;\;
    A_s=\displaystyle{\sum_i}\rho_s(b_1^i,c_1^i)d\rho_s(b_2^i,c_2^i)\\[2mm]
\Phi&=\displaystyle{\sum_i}\rho(a_1^i,b_1^i,c_1^i)[\cM,\rho(a_2^i,b_2^i,c_2^i)]=
     \left(
     \ba{cc}
     \left(
     \ba{cc}
     0&\rho_1(\varphi)M_1\\
     {M_1}^{\dag}{\rho_1}^{\dag}(\varphi)&0\\
     \ea
     \right)&0\\
     0&0\\
     \ea
     \right)\\[2mm]
H&=\Phi+\cM_1
  =\left(
   \ba{cc}
   \left(
   \ba{cc}
   0&\rho_1(h)M_1\\
   {M_1}^{\dag}{\rho_1}^{\dag}(h)&0\\
   \ea
   \right)&0\\
   0&0\\
   \ea
   \right)
\label{eqn:4-10}
\ea\ee     
with
\be
J\left(
 \ba{cc}
 p&0\\
 0&q\\
 \ea
 \right)J^{\dag}=\left(
 \ba{cc}
 q^{*}&0\\
 0&p^{*}\\
 \ea
 \right)
\label{eqn:4-11}
\ee     
Here $\varphi$ is defined by Eq.(\ref{eqn:3-6}), $h$ by Eq.(\ref{eqn:3-7})
and $\rho_1(a), a\in\bH\;$ by Eq.(\ref{eqn:4-8}).
Note that in the basis (\ref{eqn:4-6}) we have the following mass matrix:
\be\ba{cl}
M_1&=\left(
    \ba{cc}
    M_q\otimes 1_3&0\\
    0&M_l\\
    \ea
    \right),\;\;
    M_q=\left(
        \ba{cc}
        M_u&0\\
        0&M_d\\
        \ea
        \right),\;\;\;
     M_l=\left(
        \ba{cc}
        M_\nu&0\\
        0&M_e\\
        \ea
        \right)\\[2mm]
&M_u={\rm diag}(m_u,m_c,m_t,\cdots),\;\;M_d=U_q{\rm diag}(m_d,m_s,m_b,\cdots)\\[2mm]                 &M_\nu={\rm diag}(m_{\nu_e},m_{\nu_\mu},m_{\nu_\tau},\cdots),
       \;\;M_e=U_l{\rm diag}(m_e,m_\mu,m_\tau,\cdots)                               
\label{eqn:4-12}
\ea\ee     
with
obvious notations for fermion masses. 
KM matrices $U_f,f=l,q$ disappear due to the corresponding factors 
in Eq.(\ref{eqn:4-6}). Recall that
$M_f, f=q,l$ are assumed to be $2N_g\times 2N_g$ matrices.
\\
\ind
In Eq.(\ref{eqn:4-9}) we have two generalized gauge potentials, 
\bA$\;$ and $J\bA J^{\dag}$.
They correspond to two factors, $U_1$ and $U_2$ in Eq.(\ref{eqn:4-2}).
In fact, we see from the definition (\ref{eqn:4-10}) and (\ref{eqn:4-11})
\be\ba{cl}
^g\bA&=U_1\bA {U_1}^{\dag}+U_1\bd {U_1}^{\dag},\;\;U_2\bA{U_2}^{\dag}=\bA\\[2mm]
J\,^g\!\!\bA J^{\dag}&=U_2(J\bA J^{\dag}){U_2}^{\dag}+U_2\bd {U_2}^{\dag},\;\;
U_1(J\bA J^{\dag}){U_1}^{\dag}=J\bA J^{\dag}\\[2mm]
^g\bA+J\,^g\!\!\bA J^{\dag}&=U(\bA+J\bA J^{\dag})U^{\dag}+U\bd U^{\dag}
\label{eqn:4-13}
\ea\ee
where we have used the relation $U_1(U_2\bd {U_2}^{\dag}){U_1}^{\dag}=
U_2\bd {U_2}^{\dag}$. Hence the fermionic lagrangian (\ref{eqn:4-9})
is gauge invariant under (\ref{eqn:4-2}). What remains to be proved
is that the unimodularity condition (\ref{eqn:4-5})
determines the correct hypercharge assignment\cite{2),3),4),10)}.
To see what is happening in the theory,
we set
\be\ba{cl}
A_w&=\left(
     \ba{cc}
     A_w^L&0\\
     0&A_w^R\\
     \ea
     \right),\;\;\;
     A_s=\left(
     \ba{cc}
     A_s^L&0\\
     0&A_s^R\\
     \ea
     \right)\\[2mm]
&A_w^L=\left(
     \ba{cc}
     W\otimes 1_3\otimes 1_{N_g}&0\\
     0&W\otimes 1_{N_g}\\
     \ea
     \right)\\[2mm]
&A_w^R=W\to\left(
           \ba{cc}
           B^{*}&0\\
           0&B\\
           \ea
           \right)\;{\rm in}\;A_w^L\\[2mm]
&A_s^L=\left(
       \ba{cc}
       1_2\otimes G^{*}\otimes 1_{N_g}&0\\
       0&\left(
       \ba{cc}
       B^{*}&0\\
       0&B^{*}\\
       \ea
       \right)\otimes 1_{N_g}\\
       \ea
       \right)=A_s^R
\label{eqn:4-14}
\ea\ee
where 
\be
W=\displaystyle{\sum_i}a_1^ida_2^i,\;\;\;
B^{*}=\displaystyle{\sum_i}b_1^idb_2^i,\;\;\;
G=\displaystyle{\sum_i}c_1^idc_2^i
\label{eqn:4-15}
\ee
are flavor, Abelian and ''color'' gauge fields.
The last still contains Abelian part
\be
G=G'+\displaystyle{{1\over 3}}1_3{\rm tr}G,\;\;\;{\rm tr}G'=0
\label{eqn:4-16}
\ee
Now the unimodularity condition (\ref{eqn:4-5}) means that tr($A_w+A_s^{*}$)=0.
Noting that $W$ and $G'$ are traceless and gauge fields are chosen as antihermitian,
we get at
\be
B+{\rm tr}G=0
\label{eqn:4-17}
\ee
indicating that there is only one Abelian gauge field, which we take to be 
$U(1)_Y$ gauge field $B$.
If $B$ couples to lepton doublet with strength proportional to $-y$,
we are led to the following hypercharge assignment
\be\ba{cl}
&Y_L^l=\left(
       \ba{cc}
       -y&0\\
       0&-y\\
       \ea
       \right),\;\;\;\;
Y_L^q=\left(
       \ba{cc}
       \displaystyle{{1\over 3}}y&0\\
       0&\displaystyle{{1\over 3}}y\\
       \ea
       \right)\otimes 1_3\\[2mm]
&Y_R^l=\left(
      \ba{cc}
      0&0\\
      0&-2y\\
      \ea
      \right),\;\;\;\;
Y_R^q=\left(
       \ba{cc}
       y+\displaystyle{{1\over 3}}y&0\\
       0&-y+\displaystyle{{1\over 3}}y\\
       \ea
       \right)\otimes 1_3
\label{eqn:4-18}
\ea\ee
By rescaling
\be
B\to -\displaystyle{{ig'\over 2}}B,\;\;\;\;
W\to -\displaystyle{{ig\over 2}}W,\;\;\;\;
G'\to -\displaystyle{{ig_s\over 2}}G'
\label{eqn:4-19}
\ee
and putting $y=+1$ we are able to recover the fermionic 
lagrangian of the standard model
from Eq.(\ref{eqn:4-9}) with normalized Higgs field.\\
\ind
Next we turn to the bosonic sector where we make use of only \bA$\;$
in Eq.(\ref{eqn:4-10})
to define the field strength \bG.
The formula (\ref{eqn:2-27}) gives 
\be
\cL_B=\cL_{YM}+\cL_H^{K.E.}-V
\label{eqn:4-20}
\ee
where, putting
\be
z^2=\left(
    \ba{cc}
    z_p^2&0\\
    0&z_a^2\\
    \ea
    \right),\;\;\;\;
    \left\{
    \ba{l}
    z_p^2
    =\left(
          \ba{cccc}
          z_1^21_{6N_g}& & &\bigzerou\\
          &{z_1'}^21_{2N_g}& &\\
          & &z_2^21_{6N_g}&\\
          \bigzerol& & &{z_2'}^21_{2N_g}\\
          \ea
          \right)\\
    z_a^2
    =\left(
          \ba{cccc}
          z_3^21_{6N_g}& & &\bigzerou\\
          &{z_3'}^21_{2N_g}& &\\
          & &z_4^21_{6N_g}&\\
          \bigzerol& & &{z_4'}^21_{2N_g}\\
          \ea
          \right)\\
    \ea
    \right.
\label{eqn:4-21}
\ee
with the conditions
\be\ba{cl}
&(3z_1^2+{z_1'}^2)\displaystyle{{g^2\over 4}}N_g=1\\[2mm]
&(3z_2^2+{z_2'}^2)\displaystyle{{{g'}^2\over 2}}N_g+
(z_3^2+z_4^2+3({z_3'}^2+{z_4'}^2))\displaystyle{{{g'}^2\over 6}}N_g=2\\[2mm]
&(z_3^2+z_4^2)\displaystyle{{{g_s}^2\over 2}}N_g=1
\label{eqn:4-21n}
\ea\ee
Yang-Mills term becomes
\be
\cL_{\rm YM}=-\displaystyle{{1\over 8}}{\rm tr}(W_{\mu\nu}W^{\mu\nu}
                                                + {G'}_{\mu\nu}{G'}^{\mu\nu})
        -\displaystyle{{1\over 4}}B_{\mu\nu}B^{\mu\nu}
\label{eqn:4-22}
\ee
in terms of $SU(2)_L, SU(3), U(1)_Y$ gauge field strengths,
$W_{\mu\nu}, {G'}_{\mu\nu}, B_{\mu\nu}$, respectively,
and Higgs kinetic energy term is given by
\be
\cL_H^{K.E.}=L(D^\mu\phi)^{\dag}(D_\mu\phi)
\label{eqn:4-23}
\ee
with
\be
L=\displaystyle{{1\over 4}}
 {\rm tr}_g[3(z_1^2+z_2^2)(M_uM_u^{\dag}+M_dM_d^{\dag})
  +({z_1'}^2+{z_2'}^2)(M_\nu M_{\nu}^{\dag}+M_eM_e^{\dag})]        
\label{eqn:4-24}
\ee
Note that $\phi$ in (\ref{eqn:4-23}) is normalized such that
$\langle\phi\rangle=\left(
                    \ba{cl}
                    0\\
                    1\\
                    \ea
                    \right)$.
On the other hand, the potential $V$ still contains auxiliary fields to be eliminated
to take the following form
\be
V=K(\phi^{\dag}\phi-1)^2
\label{eqn:4-25}
\ee
with 
\be\ba{cl}
K=\displaystyle{{1\over 4}}
(&3z_2^2\displaystyle{\sum_{q=u,d}[{\rm tr}_g
(M_qM_q^{\dag})^2-\displaystyle{1\over N_g}}({\rm tr}M_qM_q^{\dag})^2]\\[2mm]
&+{z_2'}^2\displaystyle{\sum_{l=\nu,e}[{\rm tr}_g
(M_lM_l^{\dag})^2-\displaystyle{1\over N_g}}({\rm tr}M_lM_l^{\dag})^2])
\label{eqn:4-26}
\ea\ee
\\
\ind
The final form of the bosonic 
lagrangian after rescaling  $\phi\to 
\displaystyle{{1 \over \sqrt{L}}}\phi$ reads 
\be
\cL_B=-\displaystyle{{1\over 8}}{\rm tr}(W^2+{G'}^2)-
\displaystyle{{1\over 4}}B^2
+(D^\mu\phi)^{\dag}(D_\mu\phi)-\displaystyle{{\lambda\over 4}}(\phi^{\dag}\phi-
\displaystyle{{v^2\over 2}})^2
\label{eqn:4-27}
\ee
$W^2=W_{\mu\nu}W^{\mu\nu}$ and so on in Yang-Mills sector,
with 
\be
\lambda=\displaystyle{{4K\over L^2}},\;\;\;\;v^2=2L
\label{eqn:4-28}
\ee
The present reconstruction of the standard model
requires $N_g>1$ in order to have $\lambda>0$.
That the restriction $N_g>1$ is obtained in NCG
was first observed by Connes-Lott\cite{1)}.
It should be noted that $X$ term in Eq.(\ref{eqn:2-25})
is necessary to deduce this conclusion,
though both $X$ and the rest in  Eq.(\ref{eqn:2-25})
are separately gauge covariant.
See the footnote on p.12 as regards $N=2$ super Yang-Mills theory.
Therefore, our conclusion $N_g>1$ depends on the assumption that
the present formalism works also for $N=2$ super Yang-Mills theory.
For instance,
we never come across the same connclusion
if we modify $\bG\to{\bG}'=\bF+{\bF_0}'$,
where ${\bF_0}'=\displaystyle{\sum_i}\langle d_\chi a^i, d_\chi b^i\rangle
+\langle \Phi\chi, \Phi\chi \rangle$.
However, this modification spoils the previous success in reconstructing
$N=2$ super Yang-Mills theory in contrast with NCG consideration\cite{14)}.
The detailed derivation of Eqs.(\ref{eqn:4-24})
and (\ref{eqn:4-26})
will be postponed to the Appendix.
%%%%%%%%%%%%%%%%%%%%%%%%%%%%%%%%%%%%%%%%%%%%%%%%%%%%%%%%%%%%%%%%%%%%%%%%%%%%%%%
\section{Discussions}%Section5
%%%%%%%%%%%%%%%%%%%%%%%%%%%%%%%%%%%%%%%%%%%%%%%%%%%%%%%%%%%%%%%%%%%%%%%%%%%%%%
\ind
Using the relation $m_W^2=\displaystyle{{1\over 4}}g^2v^2=
\displaystyle{{1\over 2}}g^2L$ (see Eq.(\ref{eqn:4-28}))
and estimating $L$ by top quark mass dominance for $N_g=3$ we find
$$
m_W=m_t\sqrt{\displaystyle{{z_1^2+z_2^2\over 2(3z_1^2+{z_1'}^2)}}}
$$
If $z_1^2=z_2^2$, it follows\cite{17)} that
$$
m_t\geq \sqrt{3}m_W\simeq 139\,{\rm GeV}
$$
Hence, we are not doing so bad business, although, in fact, we get 
no constraints among the parameters due to the most general form of the
matrix $z^2$ in Eq.(\ref{eqn:4-22}).\\
\ind
Nevertheless, one may assume that NCG-based bosonic lagrangian
with $z^2$ proportional to the unit matrix $1_{32N_g}$
should become exact at some large energy scale $\Lambda$
for unknown reason.
Then at $\Lambda$ we get GUT relations
$g_s^2=g^2={5\over 3}{g'}^2$
and the top mass dominance even at such large energy scale
would lead to the mass relations 
$m_t=2m_W$ and $m_H=\sqrt{2}m_t$\cite{1),11)}\footnote{If we
introduce the parameter $\kappa^2$ in the expression for
$V$ of Eq.(\ref{eqn:2-27}),
the mass relation becomes $m_H=\sqrt{2}\kappa m_t$,
which makes it unable to predict Higgs mass by means of 
renormalization group equations.}.
We shall not pursue quantitative analysis\cite{17),21)} in terms of
renormalization group equations of these or similar constraints
in this paper.\\
\ind
\`A la Connes we are led to the rep (\ref{eqn:4-8}) by experiment if $N_g=3$. 
We do not yet
understand, however, what determines the rep (\ref{eqn:4-8})\footnote{We may argue that
the rep (\ref{eqn:4-8}) is determined by the requirements that
1) left-handed fermions be doublets, 2) right-handed fermions singlets,
3) color nonchiral, 4) $u_R$ and $d_R$ have different Abelian charges, 
5) $\nu_R$ be neutral under $U(3)\times SU(2)\times U(1)$ in each generation.
It should be noted, however, that these requirements are all
guided by experiment.}, 
including the precise
value of $N_g$ except for the theoretical restriction $N_g>1$.
Nonetheless the concept of generation is introduced into the theory
in a very natural way. And this is accompanied by
promoting Higgs field to a kind of gauge field as in more rigorous
treatment of Connes NCG. In this sense the present approach,
or more general approach of NCG offers a new insight into the
Yang-Mills-Higgs theory.
It should also be mentioned\cite{17),21)} that
Higgs mass is estimated to be below or
around the weak scale $v=$247 GeV upon quantum corrections.\\
\ind
Some authors\cite{18)} pointed out    
that the fact that the correct hypercharge assignment (\ref{eqn:4-18})
comes only from Connes unimodularity condition (\ref{eqn:4-5})
indicates that NCG-based standard model should be unified with gravity at large
energy scale\cite{22)} because Eq.(\ref{eqn:4-15}) or (\ref{eqn:4-18})
assures the absence of 
graviton-graviton-$U(1)$ anomaly, tr$Y$=0.
Our next task is to construct standard-model-gravity-coupled
model based on the present formalism.
\appendix
%%%%%%%%%%%%%%%%%%%%%%%%%%%%%%%%%%%%%%%%%%%%%%%%%%%%%%%%%%%%%%%%%%%%%%%%%%%%%%%
\newpage
\section{Appendix}%Appendix A
%%%%%%%%%%%%%%%%%%%%%%%%%%%%%%%%%%%%%%%%%%%%%%%%%%%%%%%%%%%%%%%%%%%%%%%%%%%%%%%%
Let us first derive Higgs kinetic energy term (\ref{eqn:4-23}).
By definition we have
\be
\cL_H^{K.E.}=\displaystyle{{1\over 4}}{\rm tr}z^2(D_\mu H)^{\dag}(D^\mu H)
\label{eqn:A-1}
\ee
where
\be
D_\mu H=
\left(
\ba{cc}
 \left(
 \ba{cc}
 0&D_\mu\rho_1(h)M_1\\
 M_1^{\dag}(D_\mu\rho_1(h))^{\dag}&0\\
 \ea
 \right)&0\\
 0&0\\
 \ea
 \right)
\label{eqn:A-2}
\ee
with
\be
D_\mu\rho_1(h)=(D_\mu{\tilde\phi},D_\mu\phi)\otimes 1_{4N_g},\;\;\;
\left\{
\ba{l}
D_\mu\phi=(\partial_\mu-\displaystyle{{ig\over 2}}W_\mu-
\displaystyle{{ig'\over 2}}1_2B_\mu)\phi\\
D_\mu{\tilde\phi}=(\partial_\mu-\displaystyle{{ig\over 2}}W_\mu+
\displaystyle{{ig'\over 2}}1_2B_\mu){\tilde\phi}\\
\ea
\right.
\label{eqn:A-3}
\ee
We then compute (\ref{eqn:A-1}) using the relation
\be\ba{cl}
&{\rm tr}(D^\mu{\tilde\phi},D^\mu\phi)(M_1M_1^{\dag})
\left(
\ba{cl}
(D_\mu{\tilde\phi})^{\dag}\\
(D_\mu\phi)^{\dag}\\
\ea
\right)
={\rm tr}M_1^{\dag}\left(
\ba{cl}
(D_\mu{\tilde\phi})^{\dag}\\
(D_\mu\phi)^{\dag}\\
\ea
\right)(D^\mu{\tilde\phi},D^\mu\phi)M_1\\[2mm]
&=
{\rm tr}_g[3(M_uM_u^{\dag}+M_dM_d^{\dag})+(M_\nu M_{\nu}^{\dag}+M_eM_e^{\dag})]
(D^\mu\phi)^{\dag}(D_\mu\phi)
\label{eqn:A-4}
\ea
\ee
and (\ref{eqn:4-21}) to obtain Eqs.(\ref{eqn:4-23}) and (\ref{eqn:4-24}).\\
\ind
Next we evaluate Eq.(\ref{eqn:4-26}).
According to Eq.(\ref{eqn:2-25}) we have
\be\ba{cl}
X&=\left(
  \ba{cc}
  X_1&0\\
  0&X_2\\
  \ea
  \right)
  =\left(
   \ba{cc}
     \left(
     \ba{cc}
     x_1&0\\
     0&y_1\\
     \ea
     \right)&0\\
     0&\left(
       \ba{cc}
       x_2&0\\
       0&y_2\\
       \ea
       \right)\\
    \ea
    \right)\\[2mm]
H^2&=\left(
    \ba{cc}
      \left(
      \ba{cc}
      \Theta&0\\
      0&\Theta'\\
      \ea
      \right)&0\\
      0&0\\
    \ea
    \right)\\[2mm]
M^2&=\left(
    \ba{cc}
      \left(
      \ba{cc}
      M_1M_1^{\dag}&0\\
      0&M_1^{\dag}M_1\\
      \ea
      \right)&0\\
      0&0\\
    \ea
    \right)\\[2mm]
Y&=\left(
    \ba{cc}
      \left(
      \ba{cc}
      Y_1&0\\
      0&Y_2\\
      \ea
      \right)&0\\
      0&0\\
    \ea
    \right)
\label{eqn:A-5}
\ea
\ee
with
\be\ba{cl}
\Theta&=\rho_1(h)M_1M_1^{\dag}\rho_1^{\dag}(h),\;\;
\Theta'=M_1^{\dag}\rho_1^{\dag}(h)\rho_1(h)M_1\\[2mm]
Y_1&=\displaystyle{\sum_i}\rho_1(a_1^i,b_1^i)[M_1M_1^{\dag},\rho_1(a_2^i,b_2^i)],\;\;
Y_2=0
\label{eqn:A-6}
\ea
\ee
It is now easy to eliminate the auxiliary fields, $x_1+Y_1, x_2, y_2$ 
from the expression for the potential $V$
and determine $y_1$ in the quark and lepton sectors 
from the equation of motion:
%\end{document}
\be\ba{cl}
y_1^q&=\displaystyle{{1\over N_g}}(\phi^{\dag}\phi-1)
       \left(
       \ba{cc}
       {\rm tr}_gM_u^{\dag}M_u&0\\
       0&{\rm tr}_gM_d^{\dag}M_d\\
       \ea
       \right)\otimes 1_3 \otimes 1_{N_g}\\[2mm]
y_1^l&=\displaystyle{{1\over N_g}}(\phi^{\dag}\phi-1)
       \left(
       \ba{cc}
       {\rm tr}_gM_\nu^{\dag} M_\nu&0\\
       0&{\rm tr}_gM_e^{\dag}M_e\\
       \ea
       \right)\otimes 1_{N_g}
\label{eqn:A-7}
\ea
\ee
Consequently, we arrive at the final form of the Higgs potential (\ref{eqn:4-25})
with the coefficient (\ref{eqn:4-26}).

\end{document}